\documentclass[times]{aa520}
\usepackage{psfig}
\usepackage{natbib}
\bibpunct{(}{)}{;}{a}{}{,}

\begin{document}

\def\ha{{\rm rh}}
\def\eps{\varepsilon}
\def\aap{A\&A}
\def\apj{ApJ}
\def\apjl{ApJL}
\def\mnras{MNRAS}
\def\aj{AJ}
\def\nat{Nature}
\def\azh{Astronomicheskii Zhurnal}
\def\aaps{A\&A Supp.}
\def\figsize{0.5}

\title{The Radio Luminosity Function of Cluster Radio Halos}

\titlerunning{The RLF of Cluster Radio Halos}

\author{Torsten A. En{\ss}lin$^1$ \and Huub R{\"o}ttgering$^{2}$}

\authorrunning{T. A. En{\ss}lin \and H. R{\"o}ttgering}

\institute{$^1$ Max-Planck-Institut f\"{u}r Astrophysik,
Karl-Schwarzschild-Str.1, Postfach 1317, 85741 Garching, Germany\\
$^2$ Sterrewacht, Oort Gebouw, P.O. Box 9513, 2300 RA Leiden, The Netherlands}

\date{today} 

\abstract{A significant fraction of galaxy clusters exhibits cluster
wide radio halos. We give a simple prediction of the local and higher
redshift radio halo luminosity function (RHLF) on the basis of (i) an
observed and a theoretical X-ray cluster luminosity function (XCLF)
(ii) the observed radio--X-ray luminosity correlation (RXLC) of galaxy
clusters with radio halos (iii) an assumed fraction of $f_\ha \approx
\frac{1}{3}$ galaxy clusters to have radio halos as supported by
observations.  We then find 300-700 radio halos with $S_{\rm 1.4 GHz}>
1\,$mJy, and $10^5-10^6$ radio halos with $S_{\rm 1.4 GHz}>
1\,\mu$Jy should be visible on the sky.  14\% of the $S_{\rm 1.4 GHz}>
1\,$mJy and 56\% of the $S_{\rm 1.4 GHz}> 1\,\mu$Jy halos are located
at $z>0.3$.
Subsequently, we give more realistic predictions taking into account
(iv) a refined estimate of the radio halo fraction as a function of
redshift and cluster mass, and (v) a decrease in intrinsic radio halo
luminosity with redshift due to increased inverse Compton electron
energy losses on the Cosmic Microwave Background (CMB). We find that
this reduces the radio halo counts from the simple prediction by only
30 $\%$ totally, but the high redshift ($z>0.3$) counts are more
strongly reduced by 50-70\%. 
These calculations show that the new generation of sensitive radio
telescopes like LOFAR, ATA, EVLA, SKA and the already operating
GMRT should be able to detect large numbers of radio halos and will
provide unique information for studies of galaxy cluster merger rates
and associated non-thermal processes.
\keywords{X-rays: galaxies: clusters -- Radiation mechanism:
non-thermal -- Radio continuum: general -- Intergalactic medium --
Galaxies: cluster: general}}  \maketitle

\section{Introduction\label{sec:intro}}
\subsection{Cluster radio halos\label{sec:crh}}

The X-ray emitting intra-cluster medium (ICM) of a significant
fraction of galaxy clusters also exhibits cluster wide radio emission,
the so called {\it cluster radio halos} \citep[ for recent
samples]{1996IAUS..175..333F, 1999NewA....4..141G,
2001ApJ...548..639K, 2000NewA....5..335G}. Cluster radio halos are
central, extended over cluster-scales, unpolarised, and steep spectrum
radio sources not associated with individual galaxies. It is
recognised that radio halos appear in clusters which have recently
undergone a major merger event
\citep{1993MNRAS.263...31T,2001ApJ...553L..15B}.

Whereas the cluster X-ray emission is due to thermal electrons with
energies of several keV, the emission of the radio halo is due to
synchrotron radiation of relativistic electrons with energies of $\sim
10$ GeV in $\sim\mu$G magnetic fields.  The spatial distribution of
the radio emission often seems to follow closely (and nearly linearly)
on the large scale the distribution of the X-ray emission
\citep{2001A&A...369..441G}. In a few cases, where a cluster merger is
in its early stage, detailed observations indicate that the radio
halos seem to be spatially restricted to hot merger-shocked regions
\citep{2001ApJ...559..785K,2001ApJ...563...95M}.

\begin{figure}[t]
\begin{center}
\psfig{figure=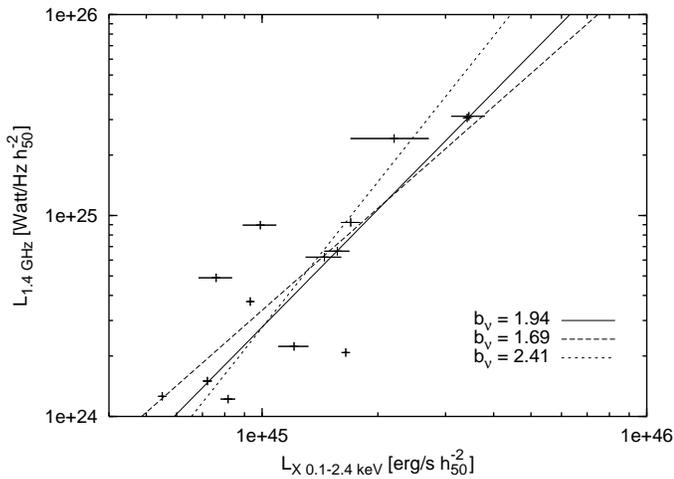,width=\figsize\textwidth,angle=0}
\end{center}
\vspace{-0.7cm}\caption[]{\label{fig:lnulX} X-ray and radio luminosity
of cluster of galaxies with radio halos. Data is from
\cite{Feretti.Pune99} and \cite{2001A&A...376..803G} and the
correlation power-laws are given in Sect.~\ref{sec:corr}.  }
\end{figure}

The similarity of X-ray and radio morphologies of radio halo galaxy
clusters indicates a connection between the energetics of the
non-thermal component (magnetic fields and relativistic electrons) and
the thermal ICM gas.  This is also supported by the strong correlation
of radio halo luminosity and the host cluster X-ray luminosity
\citep[the RXLC,][ also see Fig.~\ref{fig:lnulX}]{2000ApJ...544..686L,
Feretti.Pune99}.  Since most of the thermal cluster gas was heated in
cluster accretion and cluster merger shock waves
\citep[e.g.][]{1972A&A....20..189S, 1998ApJ...502..518Q,
2000ApJ...542..608M} one would suspect that also the relativistic
electrons received their energy from these shocks.

The radiative lifetime of the radio emitting electrons is of the order
of 0.1 Gyr \citep[e.g.][]{1977ApJ...212....1J}. This is short compared to the
shock crossing time in merger events, which is of the order of 1
Gyr. If the electrons were accelerated in the shock waves, and just
are cooling behind them, the radio emission would not follow the X-ray
emission, as observed in late stage merger clusters, but should be
more patchy and only located close to the shock waves\footnote{Such
patches of radio emission, the so called {\it cluster radio relics},
are indeed observed in merging clusters. They are interpreted to be
either emission from shock accelerated ICM electrons
\citep{1998AA...332..395E, 1999ApJ...518..603R, 2001ApJ...562..233M}
or from shock revived fossil radio cocoons \citep{2001A&A...366...26E,
2002MNRAS.331.1011E}.}. In order to have a radio halo in the post
shock region, which lasts sufficiently long to explain the X-ray
emission like morphology of radio halos in later stage mergers, some
fraction of the shock released energy has to be stored in some form
and later given to the relativistic radio emitting electron
population.

\subsection{Halo formation scenarios}

A suggestion for such an energy storing agent is turbulence within the
cluster which may re-accelerate a low energy relativistic electron
population against radiative losses \citep[][ and many
others]{1977ApJ...212....1J}. Such a primary electron model seems
to be favoured observationally by spectral index steepening towards
higher frequencies as observed in the case of the Coma cluster radio
halo
\citep[][]{1987A&A...182...21S, 2001MNRAS.320..365B}.

Another suggestion is a shock accelerated population of relativistic
protons. Over their long lifetimes they are able to inject the
necessary radio emitting relativistic electrons by charged pion decay
after hadronic interactions with the thermal ICM nucleons \citep[][
and others]{1980ApJ...239L..93D}. Such a hadronic scenario for radio
halo formation was shown to lead naturally to a very steep RXLC
\citep{Ringberg99Colafrancesco,2000A&A...362..151D,2001ApJ...562..233M},
as observed.  Such a scenario has -- in contrast to the primary
models -- difficulties to explain a strong spectral steepening, as it
seems to be apparent in the Coma cluster \citep{2002BrunettiTaiwan}.
However, measurements of the spectral indices of faint and very
extended sources, in the presence of strong point sources, are an
observational challenge, so that the possibility of larger
uncertainties in the determined radio halo spectra can not be fully
excluded yet.  The hadronic scenario will soon become further
testable since the gamma radiation from the unavoidable neutral pion
decay should be detectable by future gamma ray telescopes like GLAST
\citep{1982AJ.....87.1266V, 1997ApJ...477..560E, 1998APh.....9..227C,
2000A&A...362..151D, 2001ApJ...559...59M}.

There are also other suggested radio halo formation scenarios: radio
halos were proposed to be superpositions of large numbers of relic
radio galaxies \citep[][ and others]{1978A&AS...34..117H}, they were
proposed to be due to rapidly diffusing electrons escaping from radio
galaxies \citep[][ and others]{1979ApJ...228..576H}, and their
relativistic electrons were proposed to result from annihilation of
neutralinos, if neutralinos are the dominant dark matter component
\citep{2001ApJ...562...24C}. Although these are interesting
possibilities, they are disfavoured by the apparent association of
radio halos with merger shock waves as discussed above.

\subsection{Scientific potential\label{sec:scipot}}

In any scenario, cluster radio halos give us deep insight into the
physics and properties of galaxy clusters. Very likely radio halos
give a unique probe of non-thermal processes accompanying energetic
cluster merger events.

Large numbers of galaxy clusters are expected to be found also at high
redshifts by future surveys: e.g. the XMM Large Scale Structure Survey
is expected to find $\sim 10^3$ galaxy clusters up to redshift one
\citep{2002A&A...390....1R}, Sunyaev-Zeldovich effect cluster
detections with the Planck satellite should find $\sim 10^4$ galaxy
clusters and the Sloan Digital Sky Survey is expected to identify
$\sim 5\cdot 10^{5}$ clusters \citep{2002A&A...388..732B}.  Using
radio halos as tracers of cluster mergers should therefore allow
detailed studies of the higher redshift cluster formation processes
and properties of the accompanying cluster merger shock waves
\citep{1998ApJ...502..518Q, 2000ApJ...542..608M}. This will be
possible due to the strongly increased sensitivity and resolution of
next generation radio telescopes (e.g. ATA, EVLA, GMRT, LOFAR,
SKA). In order to guide the design and observing strategies of
these upcoming radio telescopes predictions for the number of
observable radio halos are needed. It is the aim of this paper to
provide such predictions, to show their dependence on parameters not
yet well constrained, and to indicate their scientific potential.

\subsection{Structure of the paper\label{sec:struct}}

Our predictions are based on (i) estimates of the fraction of clusters
containing halos, (ii) the local XCLF and various forms of evolution
towards higher redshift, and (iii) the local relation between X-ray
and radio halo luminosity of clusters (RXLC).

Having the halo fraction $f_\ha$ (Sect.~\ref{sec:frac}) and the RXLC
(Sect.~\ref{sec:corr}) the observed present XCLF
(Sect.~\ref{sec:xclf}) can be translated into the local RHLF
(Sect.~\ref{sec:RHLF}). In order to have predictions for higher
redshifts, where the XCLF is not yet measured, we translate a
theoretical cluster mass function into an XCLF via a mass-X-ray
luminosity correlation (MXLC) of clusters of galaxies
(Sect.~\ref{sec:xclf}). This also allows predictions of the number
counts of cluster radio halos as a function of apparent flux density
(Sect.~\ref{sec:RHLF}). We do this for a constant halo fraction
irrespective of cluster mass and redshift, and for one which evolves as
the fraction of clusters with recent mergers
(Sect.~\ref{sec:frac}). In the latter more realistic calculations we
also include a possible dimming effect of halos due to higher
radiative losses at higher redshifts. The cluster radio halo detection
strategies and expectations are briefly discussed in
Sect. \ref{sec:diss}

Our calculations are done for a $\Lambda$CDM-Universe with $\Omega_0 =
0.3$, $\Omega_\Lambda = 0.7$, $H_0 = 50\,h_{50}\,{\rm km/s}$,
$\sigma_8 = 0.9$, and $\Gamma =0.21$.

\section{Radio halo fraction\label{sec:frac}}

\begin{figure}[t]
\begin{center}
\psfig{figure=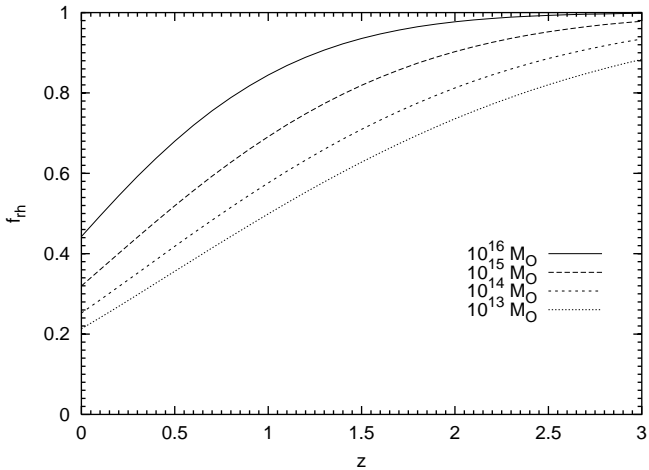,width=\figsize\textwidth,angle=0}
\end{center}
\vspace{-0.7cm}\caption[]{\label{fig:frh} Fraction of clusters which
had a strong mass increase by more than 40\% of their present mass
recently (within half a dynamical timescale $\approx 0.09/H(z)$) as a
function of their observed redshift and mass in a
$\Lambda$CDM-Universe.  }
\end{figure}

\cite{Giovannini.Pune99} find that the detection probability for a
radio halo is of the order of 0.3-0.4 for very luminous
clusters. For less luminous clusters they report a detection
probability of as low as 0.05.  Such a low detection rate can
arise naturally in a flux limited sample, even if the halo fraction is
much higher, if most of the sources are below the detection limit. In
our case, the low X-ray luminosity clusters are also expected to
contain the weakest radio halos (see Fig. \ref{fig:XCLF}), which are
therfore expected to be most likely missed in sensitivity limited
radio observations. Thus we feel that this low number of radio
halos in low X-ray luminosity clusters is likely due to a selection effect
naturally arising in searches for radio halos with luminosities
close to or below the frontier of observational feasibility.

Here, we assume that every cluster which recently grew strongly
exhibits a radio halo. Thus we implicitly assume that the radio halo
energy release is somehow delayed after the cluster merger shock
passage, as discussed in the introduction.  As a crude rule of thumb
we adopt a constant value of $f_\ha = \frac{1}{3}$ for the fraction of
clusters with a radio halo as indicated by observations of high
X-ray luminosity clusters. This number is smaller than the number of
cluster of clusters exhibiting substructure \citep[40\%-60\%,
e.g. ][]{1995ApJ...447....8M, 1999ApJ...511...65J,
2001A&A...378..408S}, but only large merger seem to produce radio
halos \citep{2001ApJ...553L..15B}.

This number can also be estimated with the help of the extended
Press-Schechter formalism\footnote{We use Eq. 2.26 in
\cite{1993MNRAS.262..627L} to estimate the conditional probability
that a cluster of given mass $M_2$ at redshift $z_2$ had a progenitor
which was more massive than $0.6 M_2$ at an earlier redshift $z_1$,
which gives us the fraction of clusters without recent strong merging
$1-f_\ha$. Contrary to a statement in \cite{1993MNRAS.262..627L},
\cite{2002MNRAS.331...98V} demonstrates that this formula gives an
accurate estimate of the extended Press-Schechter prediction of this
probability. Compared to numerical CDM simulations the agreement is
worse, but  acceptable for our purpose. We use the critical overdensity
parameter $\delta_{\rm c,0}(z)$ and the mass variance of the smoothed
density field $\sigma(M)$ in the parametrization given in
\cite{2002MNRAS.331...98V} and adopt a $\Lambda$CDM cosmology as
defined in the introduction.}, and thereby extrapolated to different
cluster mass ranges and higher redshifts. If one assumes that all
clusters which had a mass increase of more than 40\% of their final
mass within half a dynamical timescale of the final cluster (which is
approximately $\Delta t \approx 0.09/H(z)$ with $H(z)$ the Hubble
parameter at redshift z) exhibit a radio halo, one finds that
$f_{\ha}=0.32$ for present day clusters with a mass of
$10^{15}\,M_{\odot}$. The resulting halo fraction is displayed in
Fig.~\ref{fig:frh} as a function of redshift and cluster mass.

In an earlier study \cite{2001ApJ...563..660F} made a similar
calculation of the cluster merger rates. Their estimate of the
fraction of cluster radio halos is based on the radiative energy loss
timescale of the radio electrons. Since this is much shorter than the
dynamical timescale of the clusters used in our work, their fraction
of galaxy clusters exhibiting radio halos is much smaller. They find
that only 10\% of the present clusters had a major merger recently
enough (within the electron cooling time) to exhibit a radio halo. In
order to reproduce the fraction of 20\%-30\% of all present clusters,
which is indicated by observations, they require that rather weak
mergers with a mass increase of only 10\% have to be sufficient to
trigger a radio halo. In contrast to this, clusters with radio halos
exhibit signatures of much stronger merging activity than a 10\% mass
increase would produce\footnote{This can be seen by the fact that only
clusters with strong substructure have halos
\citep{2001ApJ...553L..15B}. That even a 20\% merger is insufficient
to trigger a radio halo is also demonstrated by the observations of
cluster Abell 3667. It has signatures of a recent merger since
substructure exists and two giant cluster radio relics indicate the
presence of peripheral shock waves \citep[][ and references
therein]{1997MNRAS.290..577R, 1998AA...332..395E}.  A detailed
numerical simulation of this cluster by \cite{1999ApJ...518..603R}
showed that its X-ray morphology is reproduced well by an on-axis
collision of two clusters with a mass ratio of 1:5. Although the
merger in Abell 3667 is fully developed no radio halo could be found
in the sensitive radio observations \citep{1997MNRAS.290..577R}. The
reason for this is likely that the cluster centre was not shocked in
this weak merger, as the presence of a cool core demonstrates
\citep{2001ApJ...551..160V}. On the other hand, a merger with a mass
ratio of 1:2.5 seems to be sufficiently violent to trigger a radio
halo, as the cluster Abell 754 suggests \citep{1998ApJ...493...62R,
2001ApJ...559..785K}.}.  This indicates that the relevant timescale of
halo emission after a merger should be significantly longer than the
electron cooling timescale, and is likely to be of the order of the
dynamical timescale of the merger as assumed in our work.

\section{Radio halo--X-ray luminosity correlation\label{sec:corr}}

\cite{Feretti.Pune99} compiled the properties of the presently known
cluster radio halos and cluster radio relics. In the following we use
the properties of the cluster radio halo sub-sample listed in this
work, plus the properties of the radio halo of Abell 2254, which we
take from \cite{2001A&A...376..803G}.

In Fig.~\ref{fig:lnulX}
we show the 0.1-2.4 keV X-ray and 1.4 GHz radio luminosities of the
 galaxy clusters containing radio halos. Also shown are power-law fits
of the form
\begin{equation}
\label{eq:lnulX}
L_{\nu}(L_{\rm X}) = a_{\nu}\,{10^{24}\,h_{50}^{-2}\,{\rm Watt/Hz}} \, \left(
\frac{L_X}{10^{45}\,h_{50}^{-2}\,{\rm erg/s}} \right)^{b_{\nu}}\,.
\end{equation}
We obtain the parameters $a_\nu = 3.37$ and $b_\nu = 1.69$ from linear
regression in logarithmic units with $\log(L_\nu)$ as the dependent
variable.  If $\log(L_{\rm X})$ is assumed to be the dependent
variable we get $a_\nu = 2.77$ and $b_\nu = 2.41$. A fit using errors
in both observables (assuming an uncertainty of $\Delta L_\nu/L_\nu =
0.1$) yields $a_\nu = 2.78$ and $b_\nu = 1.94$. We use all three
parameter sets to calculate the local RHLF, but favour the latter
parameters with intermediate slope since the other slopes are likely
affected by the scatter in the data.

It should be noted that the fitted RXLCs are used throughout this
paper to extrapolate halo properties for lower and higher luminosities
than yet observationally constrained. The underlying idea is that
several of the proposed scenarios for radio halo formation discussed
in the introduction predict such or similar scaling relations. At the
moment, this extrapolation is therefore only an educated guess, which
should be tested by more sensitive future observations.

There might be an additional redshift dependence of the radio halo
luminosity.  For our models with constant halo fraction we do not
assume any redshift dependence of the RXLC in order to keep the model
simple. For the more realistic scenario with evolving halo fraction we
assume $L_{\nu}(L_{\rm X},z) = L_{\nu}(L_{\rm X})\, (1+z)^{-4}$, since
for weak cluster magnetic fields ($B \le \mu$G) the ratio of
synchrotron to total (synchrotron plus inverse Compton) energy losses
is proportional to the inverse CMB photon energy density. If the
typical cluster magnetic field energy densities are comparable or even
stronger than the CMB energy density, this approach underestimates the
radio halo luminosity. Hence, by including it we give a conservative
estimate.

\section{X-ray cluster luminosity function\label{sec:xclf}}
\begin{figure}[t]
\begin{center}
\psfig{figure=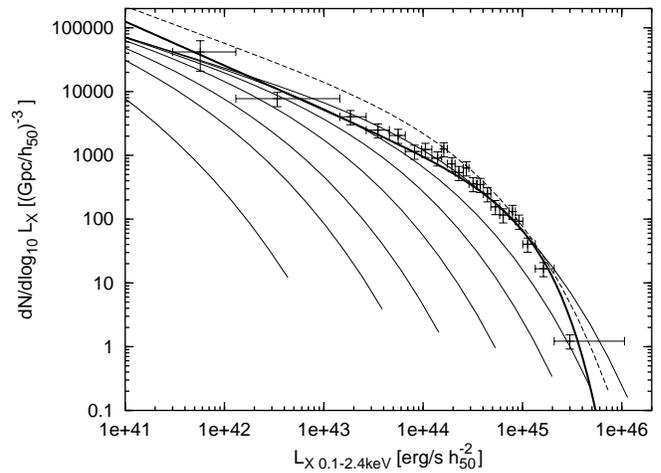,width=\figsize\textwidth,angle=0}
\end{center}
\vspace{-0.7cm}\caption[]{\label{fig:XCLF} X-ray luminosity
function. Data points and thick line are from
\cite{2002ApJ...566...93B}. The dashed line is the
\cite{2001MNRAS.321..372J} mass function translated with the help of
the \cite{2002ApJ...567..716R} MXLC. The thin solid lines are the same
mass functions translated with our adapted MXLC for redshift $z =$ 0,
0.3, 0.6, 1.0, 1.5, 2.0, and 3.0 from top to bottom. These lines end
where the range of the simulated mass functions end.}
\end{figure}

The XCLF in the ROSAT 0.1-2.4~keV band was recently estimated by
\cite{2002ApJ...566...93B}. They fit their data by
\begin{equation}
\label{eq:xclf}
\frac{dN_{\rm cl}}{dL_{\rm X}} = \frac{n_{\rm cl}}{L_{\rm X,*}}
\left( \frac{L_{\rm X}}{L_{\rm X,*}} \right)^{-\alpha_{\rm X}}
\exp\left(- \frac{L_{\rm X}}{L_{\rm X,*}} \right)\,,
\end{equation}
with $L_{\rm X,*} = 8.36\cdot 10^{44} \,h_{50}^{-2}\, {\rm erg/s}$,
$n_{\rm cl} = 107\,({\rm Gpc}/h_{50})^{-3}$, and $\alpha_{\rm X} =
1.69$ for a $\Lambda$CDM cosmology (see Fig.~\ref{fig:XCLF}).

To be able to extrapolate this locally determined XCLF to higher
redshifts, we take a two-step approach. First, we relate a halo mass
to an individual X-ray luminosity. Second, the models for the growth
of the halo masses with time then naturally give the evolution of the
XCLF.

We translate the cluster mass ($M_{200,J}$, ``J'' stands for Jenkins,
etc.)  function of \cite{2001MNRAS.321..372J} into an X-ray luminosity
function. This can be done with the empirical MXLC of
\cite{2002ApJ...567..716R}, which is based on hydrostatic cluster mass
estimates\footnote{Care has to be taken even though both works give
$M_{200}$, the mass contained in a region with an overdensity of a
factor 200, since \cite{2002ApJ...567..716R} refer to the critical
density $\varrho_c$, whereas \cite{2001MNRAS.321..372J} refer to the
cosmic mean density $\varrho_0 = \Omega_0\,\varrho_c$. We correct for
this by using $M_{200, \rm R\&B} \approx M_{200, \rm
J}\,\Omega_0^{1/2}$, which is exact for a singular isothermal sphere
and therefore acceptable for the large cluster radii involved.}:
\begin{equation}
\label{eq:LXM}
L_{\rm X} = a_{\rm X}\, 10^{45}\, h_{50}^{-2}\, {\rm erg/s}  \left(
\frac{M_{200,R\&B}}{10^{15}\,h_{50}^{-1}\, M_\odot} \right)^{b_{\rm X}}\,,
\end{equation}
where $a_{\rm X} = 0.511$ and $b_{\rm X} = 1.571$ (for their
BCSE-Bisector fit of their extended sample).  The resulting local XCLF
significantly deviates from the observed one\footnote{This may be
caused by a different set of cosmological parameters \cite[a better
agreement with a $\Lambda$CDM Universe with $\Omega_0 = 0.12$ and
$\sigma_8 = 0.96$ was found by ][]{2002ApJ...567..716R}, or by
systematic uncertainties in the underlying hydrostatic mass
estimates.} (see Fig.~\ref{fig:XCLF}), and we therefore do not use it
any further.

\begin{figure}[t]
\begin{center}
\psfig{figure=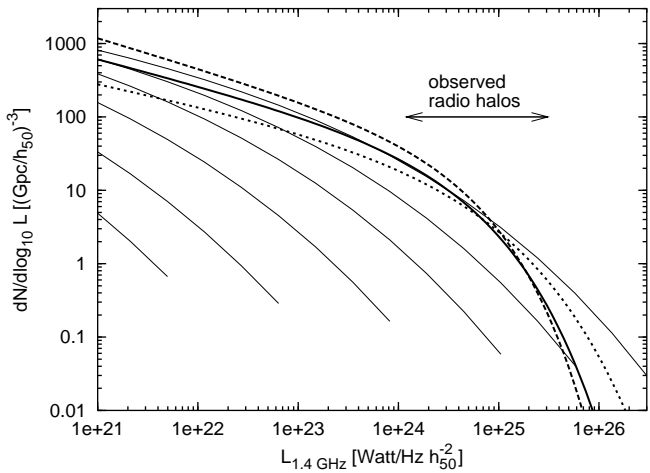,width=\figsize\textwidth,angle=0}
\end{center}
\vspace{-0.7cm}
\caption[]{\label{fig:haloRLF} Radio halo luminosity function,
derived under the assumption that a constant fraction $f_\ha = 1/3$
of all clusters contain a radio halo. The thick lines are calculated
from the observed X-ray luminosity function, which was translated
using the correlations displayed in Fig.~\ref{fig:lnulX}. The thin
solid lines are calculated from our adapted X-ray luminosity function
using the intermediate steep correlation displayed in
Fig.~\ref{fig:lnulX} for redshift $z =$ 0, 0.3, 0.6, 1.0, 1.5, 2.0,
and 3.0 from top to bottom. These lines end where the range of the
underlying simulated mass functions end.}
\end{figure}

In order to have a working model XCLF we adopt the mass function of
\cite{2001MNRAS.321..372J} and re-fitted the parameters in
Eq. \ref{eq:LXM} so that the measured local XCLF is reproduced within
the sampled range (see Fig.~\ref{fig:XCLF}). This gives $a_{\rm X}
= 0.449$ and $b_{\rm X} = 1.9$, which we adopt in the following and
denote it as our adapted MXLC (note that we insert $M_{200, \rm R\&B}
\approx M_{200, \rm J}\,\Omega_0^{1/2}$ in Eq. \ref{eq:LXM} in order
to correct for the difference in the definitions of the cluster
masses).

Our model predicts some evolution of the high luminosity end of the
XCLF at moderate redshifts ($0<z<1$). This may be in conflict with
measurements of the higher redshift XCLF, which do not reveal very
significant evolution of the XCLF in this redshift range \citep[][ for
a discussion]{1999ApJ...513L..17D}. On the other hand, the error bars
of these measurements are still quite large and could be consistent
with the amount of evolution given in our model. In order to also
cover the possible case that there is no evolution in the XRF in the
redshift range most important for the radio halo source counts, we
also present calculations in which the local XCLF is assumed to hold
at all redshifts. This gives a much larger number of radio halos. Thus
our evolving XRF model can be regarded to be conservative since it may
underpredict the number of luminous clusters in X-ray and radio.

\section{Radio halo luminosity function\label{sec:RHLF}}

\begin{figure}[t]
\begin{center}
\psfig{figure=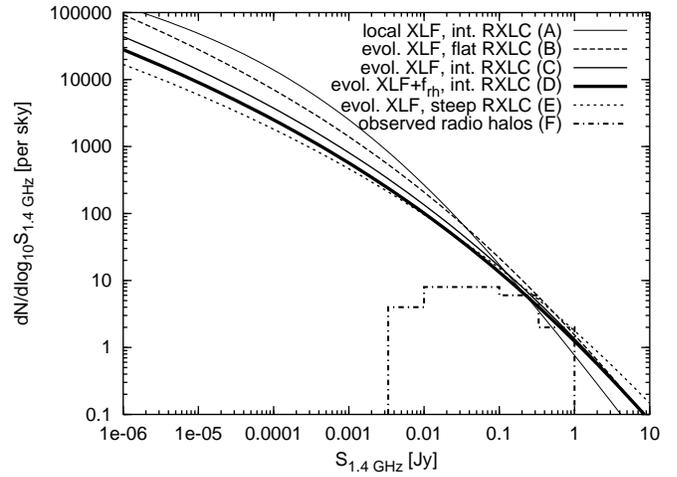,width=\figsize\textwidth,angle=0}
\end{center}
\vspace{-0.7cm}\caption[]{\label{fig:dNdS} Expected flux density
ditribution of radio halos.  All solid lines show models using the
intermediate radio--X-ray luminosity correlation. The thin solid line
(A) shows the flux density distribution if the local intermediate
radio halo luminosity function (thick solid line in
Fig.~\ref{fig:haloRLF}) is assumed to hold at all redshifts.  The
heavy solid line (D) uses the adapted X-ray luminosity function, but
assumes that the fraction of clusters with radio halos are the one
with recent mergers as displayed in Fig.~\ref{fig:frh}. In addition a
$(1+z)^{-4}$ decline in radio luminosity is assumed in that latter
model as a consequence of the higher inverse Compton energy losses on
the CMB at higher redshift. The lines B, C, and E result from our
adapted model, using the three radio--X-ray luminosity correlations
(RXLC) displayed in Fig.~\ref{fig:lnulX}. Finally, the histogram (F)
shows the flux density distribution of the cluster sample compiled by
\cite{Feretti.Pune99} (plus A2254 from \cite{2001A&A...376..803G}).}
\end{figure}

\begin{figure}[t]
\begin{center}
\psfig{figure=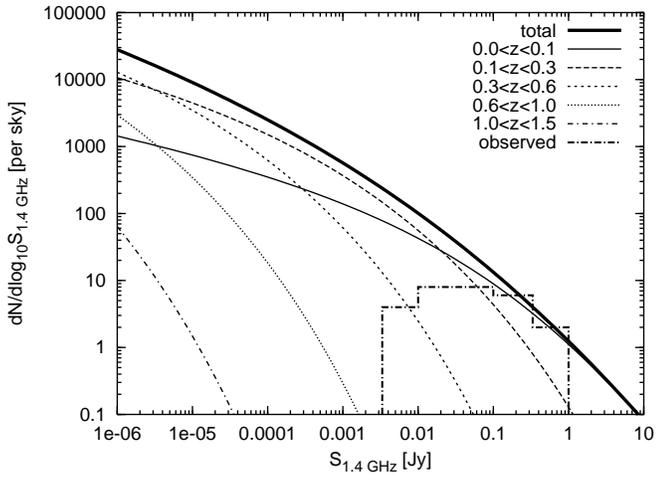,width=\figsize\textwidth,angle=0}
\end{center}
\vspace{-0.7cm}\caption[]{\label{fig:dNdSevolv} Expected flux density
distribution of radio halos for the most realistic model (see Fig.
\ref{fig:dNdS} for a comparison). The smooth curves below give the
radio halo flux density distribution for restricted redshift ranges as
indicated in the figure. The histogram shows the flux density
distribution of the cluster sample compiled by \cite{Feretti.Pune99}
(plus A2254 from \cite{2001A&A...376..803G}) which contains
clusters in the redshift range up to z= 0.55.}
\end{figure}

The local RHLF follows from Eqs. \ref{eq:lnulX} and \ref{eq:xclf}:
\begin{equation}
\label{eq:RHLFlocal}
\frac{dN_\ha}{dL_\nu} = \frac{n_\ha}{L_{\nu,*}} \, \left(
\frac{L_\nu}{L_{\nu,*}} \right)^{-\alpha_\ha} \, \exp\left(
- \left[ \frac{L_\nu}{L_{\nu,*}} \right]^{\beta_\ha} \right)\,,
\end{equation}
with $n_\ha = f_\ha\,n_{\rm cl}/b_\nu$, $L_{\nu,*} = L_{\nu}(L_{\rm X,*})$,
$\alpha_\ha = (\alpha_{\rm X}+b_{\nu} -1)/b_{\nu}$, and $\beta_\ha =
1/b_\nu$.  This is displayed for the different RXLCs in
Fig.~\ref{fig:haloRLF} together with the RHLF computed from our
adapted XCLF and the intermediate steep RXLC.

In order to be able to calculate the radio halo number counts, we fit
the adapted RHLF by a functional form like Eq.~\ref{eq:RHLFlocal}
separately for various redshifts. This allows to extrapolate to higher
radio halo luminosities and therefore to calculate the flux density
distribution by integration over the evolving radio halo
population. We assume a common radio halo spectral index of
$\alpha_\nu = 1$ for this.

The resulting flux density distribution (Fig.~\ref{fig:dNdS}) should
depend little on cosmology for larger fluxes. This is because it is
dominated at the bright end by the local RHLF, which was fixed by
observational constraints. In order to illustrate this
Fig.~\ref{fig:dNdS} also contains the flux density distribution calculated
by using the local RHLF given by Eq.~\ref{eq:RHLFlocal} for all
redshifts. Also included in this figure are more realistic
calculations including a non-constant halo fraction as displayed in
Fig.~\ref{fig:frh} and dimming at higher redshifts.  For the more
realistic scenario (evolving XCLF, evolving $f_\ha$, redshift dimming)
the contributions of different redshift ranges to the flux density
distribution is displayed in Fig.~\ref{fig:dNdSevolv}.  Further, we
have included in both figures a histogram with the observed flux density
distribution of the radio halo sample of \cite{Feretti.Pune99}
(plus A2254 from \cite{2001A&A...376..803G}). The large
discrepancy between the observed and expected flux density distribution
indicates a large incompleteness of our present knowledge of faint
cluster radio halos.

\begin{figure}[t]
\begin{center}
\psfig{figure=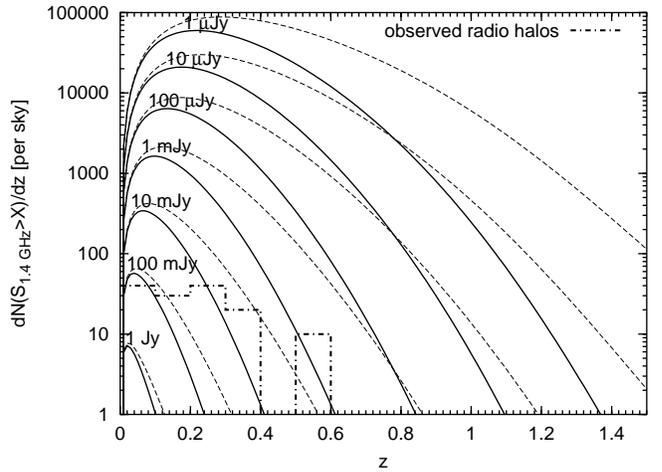,width=\figsize\textwidth,angle=0}
\end{center}
\vspace{-0.7cm}\caption[]{\label{fig:dNdz} Expected redshift
distribution of radio halos with fluxes above flux limits as indicated
in the figure. The solid lines give the most realistic model, whereas
the dashed lines do not include any radio halo dimming with
redshift. The histogram shows the differential  redshift distribution of the
radio halo cluster sample compiled by \cite{Feretti.Pune99} and
\cite{2001A&A...376..803G} (binned into bins of width $\Delta z = 0.1$).}
\end{figure}

The redshift distribution of radio halos above given flux limits is
displayed in Fig.~\ref{fig:dNdz} for the most realistic model.  A
comparison with the already observed population of radio halos shows
that the number of higher redshift clusters ($0.3 < z < 0.6$) observed
is close to the predictions. This means that either observers already
managed to find a substantial fraction of these halos or that our most
realistic model is indeed too conservative. If redshift dimming would
not occur, e.g. because cluster magnetic fields are strong and
therefore the CMB is not the main energy loss target of the electrons,
then many more higher redshift radio halos are expected as also shown
in Fig.~\ref{fig:dNdz}.

\section{Discussion\label{sec:diss}}

We estimated the cluster radio halo luminosity function and the
expected flux density distribution by translating an observed and
a theoretical X-ray cluster luminosity function with the help of the
observed cluster radio halo--X-ray luminosity correlation. A
power-law form of this correlation was used to extrapolate into the
observationally poorly constraint regime of (weak) radio halos of low
X-ray luminosity clusters. For a simple model calculation we assumed
that a fraction $f_\ha = \frac{1}{3}$ of all clusters contain radio
halos, irrespective of redshift and cluster size. We note, that if
the halo fraction for low X-ray luminosity clusters would be much
lower, which cannot be excluded with the present day data, our
predictions based on the above halo fraction would be
overestimated. In the case that the halo fraction is the same for all
cluster, but lower than assumed here, our results can simply be
scaled.

\begin{table*}[t]
\begin{center}
\begin{tabular}{|rr|rrrrrrrl|}
\hline \multicolumn{2}{|c|}{$S_{\rm 1.4 GHz,min}$} & $N^{\rm flat}_{\rm
evol.}$ & $N^{\rm int.}_{\rm evol.}$ & $N^{\rm 
steep}_{\rm evol.}$ & $N^{\rm int.}_{\rm local}$ & $N^{\rm
int.,*}_{\rm 
local} $ & $N^{\rm int.,*}_{\rm evol.}$ & $N^{\rm int.,*}_{\rm evol.}$
& $\!\!\!\!\!\!\!\!(z\!>\!0.3)$\\
\hline
$1$ & $\mu$Jy 	& 74857.9 & 36646.9 & 15388.6 & 118854.0 & 70579.6 & 23758.5 & 10784.9 & \\ 
$10$ & $\mu$Jy 	& 19784.7 & 10269.5 &  4821.5 &  36733.0 & 19686.2 &  6812.2 &  2123.7 & \\ 
$100$ & $\mu$Jy &  4308.1 &  2403.8 &  1298.1 &   8076.4 &  4247.7 &  1653.5 &   280.9 & \\ 
$1$ & $$mJy 	&   735.7 &   450.2 &   290.6 &   1143.8 &   664.7 &   326.4 &    20.5 & \\ 
$10$ & $$mJy 	&    93.3 &    64.2 &    52.0 &    100.0 &    71.0 &    50.1 &     0.6 & \\ 
$100$ & $$mJy 	&     8.4 &     6.7 &     7.1 &      5.6 &     5.0 &     5.7 &     0.0 & \\ 
$1$ & $$Jy 	&     0.5 &     0.5 &     0.7 &      0.2 &     0.2 &     0.5 &     0.0 & \\
\hline
\end{tabular}
\end{center}
\caption[]{\label{tab:NS} The number $N$ of expected radio halos on
the full sky, which are above a given flux density $S_{\rm 1.4 GHz,min}$
for the flat ($N^{\rm flat}$), the intermediate ($N^{\rm int.}$), and
the steep ($N^{\rm steep}$) radio halo--X-ray luminosity correlations
displayed in Fig.~\ref{fig:lnulX}. In addition to the model with an
evolving X-ray luminosity function ($N_{\rm evol.}$, see
Fig.~\ref{fig:dNdS}) also the radio halo number counts for a redshift
independent ($=$ local) cluster distribution are given ($N_{\rm
local}$, see Fig.~\ref{fig:dNdS}) for the intermediate RXLC. Further,
the models marked by $*$ give the expected number counts assuming that
the fraction of clusters with radio halos is not $f_\ha = \frac{1}{3}$ as
assumed in the other calculations, but is given by the fraction of
clusters which had a recent strong mass increase, as displayed in
Fig.~\ref{fig:frh}. In addition to this, it is assumed that the radio
halo luminosity of a cluster with the same mass is lower by a factor
$(1+z)^{-4}$ due to the increasing inverse Compton energy losses on the
CMB. Thus, the first three columns indicate the level of uncertainty
in these calculations due to the uncertainty in the RXLC, column 4 \& 5
give an optimistic model, and the last two columns give the most
likely estimate.}
\end{table*}

The above assumptions may be questioned, since both the higher merging
rate of clusters of galaxies and also the increased electron inverse
Compton losses at higher redshifts can modify the fraction of clusters
having radio halos. For that reasons also calculations were presented
in which we tried to take both effects into account.  If our
assumptions hold, we are able to predict the number of detectable
radio halos with upcoming sensible radio telescopes like LOFAR,
ATA, EVLA, SKA, and also the existing GMRT. Detailed numbers for
the different models can be found in Tab. \ref{tab:NS}.

The LOFAR array as an example: the point source sensitivity at 120 MHz
is expected to be 0.13 mJy within 1 hour integration time and a 4 MHz
bandwidth. A survey covering half of the sky can be accomplished in a
years timescale at this frequency and with this depth. It would find
$800-1200_{-40\%}^{+80\%}$ radio halos\footnote{The first (lower)
number result from our most realistic model, the second (higher) from
the model with constant $f_\ha$ and constant RXLC; the error range
indicates the uncertainties resulting from the possible slopes of the
RXLC; a radio halo spectral index of $\alpha_\nu = 1$, which is a
conservative assumption for this purpose, was used in the frequency
interpolation.} with a significance of 10 sigma, sufficient for
further follow up observations. Within this sample
$140-300_{-40\%}^{+80\%}$ of the radio halos are expected to have
redshifts larger than 0.3.

A more efficient strategy to find cluster radio halos would be to use
the large future cluster catalogues from SDSS, PLANCK, XMM-Newton as a
target list for deep integrations with the upcoming sensible radio
telescopes. This should allow tests of many of the hypotheses (partly
used in this work) on redshift and cluster size dependencies of the
radio halo population, helping to establish cluster radio halos as a
tool to investigate galaxy cluster formation and the non-thermal
processes accompanying it.

\begin{acknowledgements}
This work benefited from discussions with M. Bartemann,
H. B{\"o}hringer, H.  Matthis, F. Miniati, P. Schuecker, F. van den
Bosch, S.D.M. White, and from comments of an anonymous referee. We
have made use of the mass function program of A. Jenkins et
al. (2001). TAE thanks the LOFAR collaboration for the invitation and
the financial support to participate in the LOFAR workshop at the
Haystack Observatory (2001) where this work was initiated. This work
was done in the framework of the {\it The Intergalactic Medium
Research Training Network} of the European Union.

\end{acknowledgements}



\end{document}